\PassOptionsToPackage{table,dvipsnames}{xcolor}

\documentclass[sigconf,nonacm]{acmart}

\setcopyright{none}

\usepackage[utf8]{inputenc}
\usepackage[T1]{fontenc}
\usepackage{amsmath}
\usepackage{xspace}
\usepackage{enumitem}
\usepackage{pifont}
\usepackage[normalem]{ulem}
\usepackage{comment}

\usepackage{graphicx}
\usepackage{array}
\usepackage{booktabs}
\usepackage{longtable}
\usepackage{tabularx}
\usepackage{tabularray}
\usepackage{calc}
\usepackage{tikz}
\usepackage{pgfplots}
\pgfplotsset{compat=1.18}

\usepackage{epigraph}
\usepackage{framed}
\usepackage[most]{tcolorbox}

\hypersetup{hidelinks}

\definecolor{testColor}{HTML}{814AB7}
\definecolor{testBrown}{HTML}{6D3C1F}
\definecolor{white}{HTML}{FFFFFF}
\definecolor{purple5}{HTML}{7209B7}
\definecolor{purple4}{HTML}{580966}
\definecolor{purple3}{HTML}{770C8A}
\definecolor{purple2}{HTML}{9410AB}
\definecolor{purple1}{HTML}{C6259E}
\definecolor{pink}{HTML}{F83A90}
\definecolor{orange2}{HTML}{FA7543}
\definecolor{orange1}{HTML}{FC9D22}
\definecolor{yellow}{HTML}{FDC500}
\definecolor{darkblue}{RGB}{0,0,139}
\definecolor{black}{RGB}{0,0,0}
\definecolor{centralfill}{HTML}{075B73}
\definecolor{presentfill}{HTML}{B8DDE3}

\newtcolorbox{keyInsightsBox}{
    enhanced,
    opacityback=0.75,
    colback=purple1!25!white,
    width=0.48\textwidth,
    colframe=white,
    arc=6mm,
    boxsep=2pt,
    top=1pt,
    bottom=1pt
}

\newcolumntype{Q}{>{\centering\arraybackslash}m{5.5mm}}

\title{DevEx is for Humans, AX is for Agents: Measuring Context Health in Agentic Software Work}

\author{Gratiana Fu}
\affiliation{%
  \institution{DX}
  \city{Salt Lake City}
  \country{USA}
}
\author{Josh Beitler}
\affiliation{%
  \institution{DX}
  \city{Salt Lake City}
  \country{USA}
}
\author{Brian Houck}
\affiliation{%
  \institution{DX}
  \city{Salt Lake City}
  \country{USA}
}
\renewcommand{\shortauthors}{Fu et al.}

\begin{document}

% Complete abstract, including its environment.
% Paste into all_files/0.abstract.tex.
% In main.tex, replace the existing empty abstract environment with:
% \input{all_files/0.abstract}
% Place that input after \begin{document} and before \maketitle.
% Do not wrap this file in a second abstract environment.
% No new packages or bibliography entries are required.

\begin{abstract}
Coding agents succeed or fail in the context humans give them, yet most evaluations measure only output. Drawing on developer experience research, we introduce Agent Experience (AX), an instrument that scores session transcripts on four dimensions of the CAFE(S) framework: Clarity, Actionability, Fidelity, and Efficiency; Security, a gate rather than a friction dimension, is out of scope. A rubric-guided language model rates each dimension from 1 to 5 and explains what the human could have provided. We report on three studies that tested this model. In the Stack Overflow Developer Survey, respondents reporting context problems were less likely to rate AI as very useful in all seven activities examined (12--28\%). Across 882 rated sessions from a pilot sample of 1,000, Actionability received the lowest ratings (59.6\% of ratings at or below 3, versus 26.6\% for Fidelity), and ratings were lower in longer sessions.  One researcher's unblinded review of 15 randomly sampled sessions found all scores reasonable and explanations supported. In a controlled scenario, we injected context defects into one data-analysis task and ran it twelve times. Every defect condition used more tokens than the clean baseline (mean $+91\%$), and five evaluator models flagged 50--85\% of targeted dimensions and 14--36\% of non-targeted dimensions. Fidelity defects were detected most consistently and Actionability defects least. Because interventions were imperfectly isolated, these results are exploratory. AX gives teams structured, actionable evidence about context-related friction that complements output and outcome measures.
\end{abstract}

\maketitle

% This file should include \begin{abstract} and \end{abstract}.
% If it already has those lines, do not add another pair here.
\section{Introduction}
\label{sec:introduction}

To understand whether developers have the conditions to do their best work, developer experience (DevEx) research turns to the developers themselves. Structured surveys make it possible to examine aspects of work that are difficult to infer from activity records alone, including cognitive load, feedback delays, and interruptions to flow~\cite{noda2023devex}. Previous empirical work has associated these dimensions with individual, team, and organizational outcomes~\cite{forsgren2024devexaction}. This perspective also informs broader approaches to engineering measurement: SPACE treats productivity as multidimensional, while EngThrive connects the conditions of engineering work with organizational outcomes~\cite{forsgren2021space,houck2026engthrive}. Together, these approaches motivate examining the conditions under which work happens alongside the work that is produced.

Agentic development raises an analogous question: do agents have the conditions they need to work effectively? As agents take on tasks that require interpreting requirements, navigating repositories, using tools, and responding to feedback, their performance becomes a property of the interaction between their capabilities and their working environment. Measuring activity and task completion is necessary, but diagnosing impediments also requires evidence about the conditions encountered during execution. The diagnostic perspective of DevEx offers a starting point for this problem: examine what happens during the work to understand the conditions that impede effective execution.

Industry data makes this question practically consequential. In DX's State of AI Impact in Engineering Q3 2026 Report, drawing on data from more than 500 engineering organizations, AI spend per developer rose from \$63 to \$1,314 over four quarters, while the median innovation ratio---the reported proportion of engineering time devoted to new features and capabilities---increased from 57\% to 59\%~\cite[pp.~3, 12, 16]{dx2026q3}. That gap raises the question of what prevents higher execution. The report identifies context quality as one possible contributor to delivery friction and explicitly presents its proposed links to engineering outcomes as hypotheses requiring measurement~\cite[pp.~17--20]{dx2026q3}. 

For an agentic task, context can include instructions, requirements, repository contents, retrieved documents, prior interaction, and tool outputs. Research on software development has long documented the importance of obtaining information needed to perform work~\cite{ko_deline_venolia_2007}. Information-quality research likewise treats quality as multidimensional and dependent on the needs of the information consumer~\cite{wang_strong_1996}. For language models, even the position of relevant information within a context can materially affect task performance~\cite{liu2024lostmiddle}. These findings motivate examining whether the context assembled for a particular task gives the agent usable conditions for completing it.

CAFE(S) provides the conceptual foundation for this examination~\cite{houck2026cafes}. The framework describes five properties of context: Clarity, Actionability, Fidelity, Efficiency, and Security. This paper focuses on the first four, which we refer to collectively as CAFE. Clarity concerns whether information has an unambiguous interpretation; Actionability concerns whether objectives, constraints, and completion criteria support proceeding with the task; Fidelity concerns whether information is accurate, current, and consistent; and Efficiency concerns whether relevant information can be used without unnecessary context or search. Security remains outside the scope of the present instrument. CAFE(S) establishes a shared language for context quality and explicitly leaves its measurement as an opportunity for subsequent work.

We introduce \emph{Agent Experience (AX)}, a measurement instrument that operationalizes CAFE through structured assessment of agent-session evidence. In this paper, we use \emph{context health} to describe the quality of the information available to an agent for the task at hand. AX translates the four dimensions into questions and scoring criteria that an evaluator applies to a recorded interaction. In the implementation studied here, a large language model serves as the evaluator, producing dimension-specific ratings and supporting explanations. This approach draws on the use of language models as rubric-guided judges~\cite{zheng2023judge}. The analogy with developer surveys is diagnostic: AX examines the conditions encountered during work through evidence in the session record. Its scores are intended to characterize the information environment as observed in that record.

Dimension-specific measurement matters because different context deficiencies suggest different interventions. An unclear term may call for disambiguation; an underspecified objective may require a definition of done; an incorrect reference may require correction; and an unnecessarily large context may require more selective information provision. A completed task can still involve substantial effort spent compensating for such deficiencies. AX is therefore designed to complement measures of task success and resource consumption by providing a structured account of the context-related friction visible during execution.

This paper investigates AX through three complementary empirical components. First, we conduct a secondary analysis of the Stack Overflow Developer Survey to examine associations between reported context problems and perceived AI usefulness across development activities. Second, we characterize the instrument using naturalistic agent-session transcripts, examining repeated assessments, alternative prompt formulations and response formats, and different evaluator models. This component considers both the stability of recorded assessments and how ratings use the available response range. Third, we conduct an exploratory scenario study based on a real data-analysis task. We construct a baseline condition and conditions intended to introduce CAFE deficiencies, including combinations of deficiencies, and execute them using a fixed agent configuration. In each session, the researcher acting as the user attempts to clarify requirements and repair the interaction as the task proceeds. We examine recorded resource use and automated task scores alongside AX assessments of the resulting transcripts. The scenario analysis retains the ordinal ratings and uses an exploratory diagnostic rule to identify high friction or a decline from the clean reference. We examine targeted and non-targeted flagging, including responses across related dimensions, and compare five evaluator models on the same transcripts.

The study addresses three research questions:
\begin{description}
    \item[RQ1: Context and effectiveness.]
    How does context quality relate to perceived AI usefulness, token usage, and task outcomes?

    \item[RQ2: Measurement behavior.]
    How do AX ratings vary across repeated assessments, prompt configurations, and evaluator models?

    \item[RQ3: Diagnostic correspondence.]
    Can AX detect CAFE deficiencies within an agent's context?
\end{description}

Our contributions are threefold. First, we specify AX as a session-level instrument for operationalizing the four CAFE dimensions. Second, we provide an empirical characterization of its rating behavior across assessment configurations. Third, we present complementary observational and scenario evidence about context problems, comparing perceived AI usefulness, resource use, automated task scores, and dimension-specific AX assessments. This work advances CAFE from a conceptual account of context quality toward an explicit measurement practice. AX supplies the procedure through which that account can be examined in agent sessions and used to guide improvements to the conditions of agentic work.

\section{Related Work}
\label{sec:related-work}

\subsection{Developer Experience and Diagnostic Measurement}

Developer experience research provides a precedent for examining the conditions of work alongside its outcomes. In \emph{DevEx in Action}, Forsgren et al.\ assessed the measurement properties of developer-reported constructs separately from their relationships with individual, team, and organizational outcomes~\cite{forsgren2024devexaction}. EngThrive combines developer surveys with telemetry and connects outcome-oriented measures to diagnostic submetrics~\cite{houck2026engthrive}. Google's EngSat program provides another organizational example, using quarterly developer surveys as an ongoing measurement practice since 2018~\cite{dangelo2024longitudinal}. These approaches integrate structured evidence about impediments to effective work into engineering measurement. AX carries this diagnostic purpose into agentic work through assessment of recorded interactions. The survey literature informs this measurement perspective; the interpretation of AX ratings requires evidence specific to its assessment procedure and the properties being measured.

\subsection{Information Environments and Context Quality}

The importance of usable information predates current AI systems. Knowledge-management research examines how organizations create, transfer, and apply knowledge~\cite{alavi_leidner_2001}. Within software engineering, observational studies document the information developers seek about program behavior, relationships, and design rationale while performing their work~\cite{ko_deline_venolia_2007,sillito_murphy_devolder_2006}. Wang and Strong's consumer-oriented account of data quality extends beyond accuracy to include properties such as relevance, interpretability, completeness, and appropriate amounts of information~\cite{wang_strong_1996}. This provides a conceptual precedent for treating the quality of an information environment as multidimensional and relative to the task and its consumer.

Language-model studies provide evidence that the organization and delivery of information can affect task performance. Liu et al.\ show that performance on long-context tasks depends on the position of relevant information within the input~\cite{liu2024lostmiddle}. Laban et al.\ examine tasks whose requirements are revealed across multiple turns and find that the tested models often make early assumptions and fail to recover reliably as further information arrives~\cite{laban2026multiturn}. In a recent preprint, Gloaguen et al.\ find that repository-level context files generally increase inference cost without significantly improving task completion in their evaluated coding-agent settings~\cite{gloaguen2026agentsmd}. These studies motivate attention to what information is supplied, how it is organized, and how it develops during an interaction.

CAFE(S) supplies a shared vocabulary for describing the resulting context and explicitly identifies measurement as subsequent work~\cite{houck2026cafes}. AX operationalizes its first four properties as assessments of the information environment represented in a session record. This connects the broader information-quality tradition with the practical problem of identifying context-related friction during agentic work.

\subsection{Automated Assessment of Context}

Prior work has operationalized several context-related properties in retrieval-augmented generation (RAG). RAGAs evaluates context relevance, answer relevance, and faithfulness~\cite{es2024ragas}. Its context-relevance metric examines whether retrieved material is focused on the question, while faithfulness examines whether the answer's claims are supported by that material. ARES assesses these components using lightweight judges trained on synthetic examples, combined with human-annotated validation data and prediction-powered inference to estimate system-level quality with uncertainty intervals~\cite{saadfalcon2024ares}.

These approaches provide a precedent for separating assessment of retrieved context from assessment of the generated answer. Their attention to relevant, focused material overlaps with CAFE Efficiency. The meaning of \emph{faithfulness}, however, concerns the relationship between an answer and its supplied evidence. CAFE Fidelity concerns the trustworthiness of the context itself, including its accuracy and currency~\cite{houck2026cafes}. An answer can faithfully reproduce information from an outdated source. Distinguishing these assessment targets is therefore necessary when interpreting context and output scores.

A recent preprint by Bousetouane addresses the broader problem of measuring agent context~\cite{bousetouane2026context}. Its ProofAgent-Harness applies a seven-criterion rubric using multiple model jurors, with criteria including role clarity, instruction consistency, grounding sufficiency, and token efficiency. It assesses assembled context before or alongside behavioral evaluation and examines relationships between context scores and agent behavior. This is a close precedent for treating an agent's information environment as an explicit object of measurement. AX develops this direction through a CAFE-based assessment of recorded sessions, with particular emphasis on characterizing the instrument's behavior and its correspondence with deliberately introduced context deficiencies.

\subsection{Language Models as Evaluators}

Language models offer a means of translating natural-language criteria into automated assessments of open-ended outputs. Zheng et al.\ examine agreement with human preferences while documenting position, verbosity, and self-enhancement biases, as well as differences in how models respond to judge prompts~\cite{zheng2023judge}. G-Eval combines explicit criteria, generated evaluation steps, and structured scoring. Its analysis identifies concentration on particular integer scores and frequent ties as limitations of direct rating~\cite{liu2023geval}. This motivates inspecting how AX uses its response scale alongside measures of agreement. A wider score distribution is useful only insofar as it reflects meaningful differences in the assessed context.

Evaluator identity can also affect assessment. In summarization experiments, Panickssery et al.\ find that models can recognize their own generations and connect this ability to preferential evaluation~\cite{panickssery2024selfpreference}. For AX, this makes the relationship between the model performing the task and the model assessing the session an important part of the study description.

Reference judgments introduce a further consideration. Guerdan et al.\ distinguish disagreement between raters from rating indeterminacy, in which a single rater can regard several ratings as reasonable under the available criteria. Their analysis shows that forcing a single rating and aggregating such judgments can distort comparisons between judge systems~\cite{guerdan2025indeterminacy}. These findings make the evaluator, prompt, scoring criteria, and interpretation of reference labels parts of the measurement procedure. AX's characterization accordingly examines repeated assessments, prompt and response-format variations, and differences among evaluator models.

\subsection{Measurement Validity and the Contribution of AX}

Measurement theory distinguishes reproducible observations from justified interpretations of those observations. Kane organizes validation around the claims and assumptions connecting observations to intended score interpretations and uses~\cite{kane_2013}. Campbell and Fiske distinguish convergence across measurement procedures from discrimination between the attributes being measured, emphasizing that observed relationships can reflect both attributes and measurement methods~\cite{campbell_fiske_1959}.

For AX, these distinctions separate questions about repeated scoring, agreement among evaluator models, and identification of particular CAFE deficiencies. We treat the four dimensions as a profile of separately interpretable context properties. Deficiencies in conceptually distinct properties can co-occur; the relevant measurement question is whether each AX rating reflects its intended property. Evaluating this profile involves examining both the dimensions targeted by an intervention and those that were not targeted. We also distinguish the ordinal profile from the exploratory decision rules used to summarize it: a rule that identifies high friction answers a different question from one that identifies a decline relative to a clean reference. Examining both the rule and cross-dimension responses is therefore part of interpreting the instrument.

The contribution of this paper is an explicit instrument for operationalizing CAFE through agent-session evidence, accompanied by empirical characterization and exploratory scenario testing. The preceding literature motivates its diagnostic purpose, its assessment targets, and the evidence needed to interpret its scores. AX brings these elements together in a procedure for studying the conditions that support effective agentic work.

\section{Methods}
\label{sec:methods}

\subsection{Study Design and Measurement Procedure}

We examined AX through three complementary analyses. A secondary analysis of the Stack Overflow Developer Survey examined associations between reported context problems and perceived AI usefulness (RQ1). An assessment of naturalistic agent-session transcripts characterized AX ratings across prompt, response-scale, and evaluator configurations (RQ2). An exploratory scenario study introduced context deficiencies into a real data-analysis task and examined execution measures alongside AX assessments (RQ1 and RQ3). The units of analysis were survey respondents, recorded agent sessions, and scenario executions, respectively.

AX operationalizes the four CAFE dimensions as an evidence-based assessment of the context encountered during a session. Security was outside the scope of this study. An evaluator received a recorded interaction and a rubric requesting separate assessments of Clarity, Actionability, Fidelity, and Efficiency. The evaluator was a language model performing a retrospective assessment; the model that originally executed the task could be different. Although the prompt adopted the executing agent's perspective, ratings were judgments about the supplied session evidence.

The reference five-point instrument, designated Prompt~1, used dimension-specific anchors on a common scale of context-related friction. A score of 5 indicated direct progress without recovery, 4 trivial recovery, 3 bounded recovery, 2 substantial recovery, and 1 a disabling deficiency. The prompt described 3 as the normal, competent baseline and instructed evaluators to start there and adjust based on evidence. It required evidence for deductions and assessment of context as it arrived. Successful recovery did not erase an earlier deficiency. Each dimension required its own supporting evidence, while a single contextual problem could affect multiple dimensions. The assessment targeted supplied instructions and materials and the coherence of the interaction; inherent task difficulty, agent competence, and operational failures were outside its scoring scope. Alternative prompts varied aspects of this procedure as described below.

\subsection{Secondary Analysis of the Stack Overflow Developer Survey}
\label{sec:methods-so}

We analyzed the public 2026 Stack Overflow Developer Survey~\cite{stackoverflow2026} to examine the association between reported context problems and perceived AI usefulness. This secondary, cross-sectional analysis addressed RQ1 at the respondent level. The two session studies described below evaluated AX directly.

\paragraph{Data and eligibility.}
We used the multiple-selection \texttt{ContextProblems} item and the seven \texttt{AIImpact} activity items, with activity labels taken from the survey schema. We considered all respondent records, without additional restrictions by occupation, employment, or AI-use frequency. We trimmed whitespace, treated blank values and the literal marker \texttt{NA} as missing, and parsed multiple selections as sets of semicolon-separated responses.

Respondents were eligible if they selected at least one context response and provided at least one nonmissing AI activity response. ``Not applicable'' counted toward this initial eligibility check but was excluded from the activity-specific comparisons.

\paragraph{Comparison groups.}
We compared two groups:
\begin{itemize}
    \item \textbf{Reference group:} respondents selecting ``I usually have the context I need,'' with no other selections except ``Not applicable'' or ``Other, please specify.''
    \item \textbf{Context-problem group:} respondents reporting outdated information, incomplete information, or different teams having different versions of the truth. Respondents selecting both one of these barriers and ``I usually have the context I need'' were included in this group.
\end{itemize}
Respondents meeting neither definition were excluded from the comparisons. The groups therefore captured a specified subset of context problems.

\paragraph{Outcome.}
For each of seven activities---communication, design, implementation, review, debugging or refactoring, architecture exploration, and documentation---we calculated the proportion selecting exactly ``Very useful.'' Each activity and group had its own denominator, excluding missing and ``Not applicable'' responses while retaining all other recorded answers.

Respondents could contribute to several activity comparisons without contributing to all seven. Group membership counts therefore differed from the activity-specific denominators, and the seven comparisons involved overlapping sets of respondents.

\paragraph{Statistical comparisons.}
For each activity, we divided the context-problem group's ``Very useful'' proportion by the reference group's proportion. A ratio below one indicated a smaller proportion of ``Very useful'' responses in the context-problem group. We calculated 95\% confidence intervals using a normal approximation on the log-ratio scale. The figure marked comparisons whose intervals excluded one.

These comparisons used unweighted counts, without adjustment for respondent characteristics or multiple comparisons. Ratios and intervals were left undefined if either group had no eligible activity responses or no ``Very useful'' responses; no continuity correction was applied.

\subsection{Instrument Characterization Using Naturalistic Agent Sessions}
\label{sec:methods-naturalistic}

We evaluated AX using transcripts of naturally occurring agent sessions collected from participants in a pilot of DX's Agent Experience feature~\cite{beitler2026agentexperience}. The pilot assessed a limited set of session-experience dimensions; this study applied the configurations described below to assess all four CAFE dimensions. We examined score distributions, relationships among dimensions, sensitivity to prompt and response-scale design, and agreement across recorded evaluations. We did not have independent reference ratings for every session.

\paragraph{Session sample.}
We randomly selected 1,000 session transcripts from the pilot collection. The selection process assembled the message records associated with each sampled session. The retained evaluation outputs contained 968 distinct sessions, identified by the combination of instance and source session identifier. We used these identifiers to match evaluations across configurations.

Sessions without a substantive task, such as greetings or idle exchanges, could receive \texttt{not\_enough\_data\_for\_evaluation}. Individual dimensions that could not be evaluated could receive \texttt{not\_applicable}. We excluded these nonnumeric responses from the relevant numerical summaries. The baseline configuration produced numerical ratings for 882 sessions. Because eligibility varied across configurations and dimensions, each comparison used the sessions with comparable numerical ratings and its corresponding denominator.

\paragraph{Information available to evaluators.}
Evaluators received the exported user and assistant messages and accompanying metadata, including session identifiers, timestamps, and available model and token-use fields. The exports did not separately supply the original executing agents' system instructions, or structured records of tool calls and tool results. Tool output, file contents, and other external materials were visible only to the extent that they appeared in the exported messages. AX ratings therefore characterized the context and interaction evidenced in that conversation record.

\paragraph{Scoring and response formats.}
The five-point variants used the friction scale described in Section~3.1; three-point variants used the categories Clean, Friction, and Blocked. Evaluators returned dimension scores and supporting explanations in structured JSON.

The prompts included handling rules for redacted images and IDE selections, attachment references, and agentic command or skill invocations. These rules generally treated referenced material as available to the original agent unless the transcript indicated otherwise. They specified how evaluators should interpret the record, rather than how the data had been deidentified.

\paragraph{Evaluation configurations.}
Table~\ref{tab:ax-configurations} summarizes the nine recorded runs. The run notes identified the evaluator for Runs 1--4 and 7--9 as \emph{Auto (Claude Opus 5.5)}, and GPT-6 Luna and GPT-6 Sol for Runs 5 and 6, respectively. These labels identify the evaluators, which could differ from the agents that originally performed the tasks.

\begin{table}[t]
  \centering
  \small
  \caption{Recorded configurations for the naturalistic session evaluations.}
  \label{tab:ax-configurations}
  \begin{tabularx}{\columnwidth}{@{}lX@{}}
    \toprule
    Runs & Configuration \\
    \midrule
    1--3 & Five-point baseline; one question per dimension; repeated recorded evaluations. \\
    4 & Five-point baseline with question perspective changed from second person to first person. \\
    5 & First-person, five-point configuration evaluated with GPT-6 Luna. \\
    6 & First-person, five-point configuration evaluated with GPT-6 Sol. \\
    7 & Three-point configuration with explicit scoring gates and anchors. \\
    8 & Three-point configuration with anchoring rules removed. \\
    9 & Five-point configuration with two questions per dimension, aggregated into dimension scores. \\
    \bottomrule
  \end{tabularx}
\end{table}

Runs 1--3 contained identical recorded scores and comments. The retained summaries did not establish whether these came from fresh, independent model calls, so we distinguish agreement among the recorded outputs from independently demonstrated repeatability.

Runs 7--8 compared anchoring configurations. Run 5 also changed the evaluation procedure: after evaluators attempted to execute tasks embedded in the transcripts, instructions to treat transcript content as untrusted input and alternative transcript-delivery methods were introduced. The cross-model comparison therefore may reflect differences in procedure as well as evaluator choice.

\paragraph{Analysis.}
We summarized numerical score distributions and the frequency of nonnumeric evaluations. For matched sessions, we examined exact agreement, agreement within one point, absolute score differences, and rank correlations. The anchoring comparison also reported weighted kappa. The two-question configuration aggregated item ratings within each dimension.

We also examined relationships among CAFE scores, associations with session message count, and differences across task and original executing-model categories. These analyses described how ratings varied across sessions and configurations; conceptually distinct CAFE properties could co-occur within a session.

\paragraph{Manual review of AX output.}
One researcher reviewed 15 session transcripts selected uniformly at random to assess the baseline AX outputs. The reviewer read each transcript alongside AX's scores, explanations, and suggested improvements, and assessed three criteria: whether each score was within one point of the score the reviewer would have assigned; whether each explanation was supported by the transcript; and whether each suggested improvement was relevant and actionable.

\subsection{Exploratory Scenario Study with Injected Context Failures}
\label{sec:scenario-methods}

We conducted a scenario-based pilot to examine how AX assessments responded to deliberately introduced context problems during an actual agent task. A researcher enacted the human participant in repeated executions of one data-analysis scenario. The study combined authored initial context perturbations with adaptive dialogue intended to help the agent complete the task. The resulting comparison concerns sessions initiated under different intended context conditions and their subsequent recovery trajectories.

\subsubsection{Task, conditions, and execution}

The task used the 2026 Stack Overflow Developer Survey and the secondary analysis described above. The agent was asked to compare respondents reporting adequate context with respondents reporting context problems, using the proportion selecting ``Very useful'' for seven AI-assisted activities. The intended deliverable comprised activity-specific percentages, relative likelihoods, group sizes, a comparison of gaps, and significance assessments. This was one underlying task repeated under 12 conditions, with one execution per condition: an intended clean baseline; four individual perturbations, C, A, F, and E; six pairs, CA, CF, CE, AF, AE, and FE; and a condition combining all four dimensions.

Conditions were specified through four authored text components. Clarity perturbations used ambiguous group and outcome references. Actionability perturbations omitted explicit deliverable and completion criteria. Fidelity perturbations supplied plausible but false descriptions of missing values, multiple selections, and a response label. Efficiency perturbations specified broad data pointers, excessive schema material, and an unrelated request. These labels encode intended manipulations, rather than independently verified diagnoses of every resulting session.

All 12 executions used Rovo CLI with the executor recorded as \texttt{claude-opus-5-5}. One researcher supplied the human turns. Sessions used separate working directories with symbolic links to a shared data location. Follow-up responses were improvised: the researcher answered questions, clarified requirements, and attempted to recover progress toward the task. Consequently, some initially deficient context was repaired during interaction. In particular, completion criteria were supplied later in the actionability conditions.

The execution audit identified departures from the intended conditions. Corrective documentation was discoverable in a shared data location and prior analysis artifacts. Later clarifications and apologies could supply additional diagnostic cues. We preserve the original condition labels for the descriptive comparisons and report these departures when interpreting them. They preclude treating non-targeted dimensions as verified defect-free controls.

\subsubsection{AX assessment procedure}

Five evaluator models scored the 12 archived scenario transcripts: Claude Haiku~4.5, Claude Haiku~5.5, GPT-6 Luna, GPT-6.1 Terra, and Claude Opus~5.5. Each evaluator produced one set of four dimension scores per transcript. These scoring runs reassessed the existing task records; the task executions themselves were held fixed.

For the evaluations, the researcher started fresh Rovo sessions with each model selected and applied a common scoring wrapper. The wrapper embedded Prompt~1's four-dimension, five-point rubric with minor wording adaptations. Its dimension questions, scoring anchors, calibration, and evidence requirements preserved the substantive assessment procedure. The outer wrapper requested the consolidated score table and allowed the researcher to interrogate decisions afterward. The reported numerical assessments were drawn from those score tables.

\subsubsection{Measures and descriptive comparisons}

We retain the ordinal dimension ratings and derive a diagnostic flag for high friction or a decline of at least two rubric categories from the corresponding clean rating. Let $s_{jcd}$ be evaluator $j$'s score for condition $c$ and dimension $d$, and let $b_{jd}$ be that evaluator's score for CLEAN on the same dimension. The flag is
\begin{equation}
\label{eq:ax-scenario-flag}
F_{jcd}=\mathbf{1}\!\left[s_{jcd}<3\;\lor\;b_{jd}-s_{jcd}\geq2\right].
\end{equation}
The absolute criterion follows Prompt~1's high-friction and disabling categories. The relative criterion also recognizes a transition from effortless execution (5) to bounded recovery (3). We selected this rule during exploratory analysis to distinguish such changes from a one-category reduction; it was not prespecified, and a two-category decline is an operational contrast rather than a validated minimum reliable change. The same rule is applied to targeted and non-targeted dimensions in all five retained score series. We additionally report threshold comparisons using absolute scores below 3, any decline from CLEAN, and scores below 5. The latter captures any departure from the rubric's effortless category.

The \emph{targeted-dimension flag rate} is the proportion of intended targeted dimension assessments receiving a flag. The \emph{non-targeted-dimension flag rate} is the corresponding proportion among assessments of dimensions not targeted by the intended condition. Each score series contributes 48 ratings across 12 runs: 20 targeted and 28 non-targeted assessments, including CLEAN's four non-targeted dimensions. Each dimension contributes five targeted and seven non-targeted assessments. We retain CLEAN to describe flagging across the full set of runs and report its profile separately. CLEAN is also the relative reference: its comparison with itself cannot satisfy the decline criterion, although a score below 3 could satisfy the absolute criterion. Rates use counts within each evaluator, rather than averages of condition-specific percentages. Intended labels do not establish which deficiencies were actually present or absent, so these rates describe diagnostic correspondence rather than validated classification accuracy. The workbook's pre-existing AX Correctness column uses a different rule and is not used for these calculations.

To examine whether Clarity and Actionability contributed to non-targeted flagging, we conducted an exploratory examination of their paired responses. In conditions targeting Clarity but not Actionability (C, CF, CE), we counted non-targeted Actionability flags; conversely, in conditions targeting Actionability but not Clarity (A, AF, AE), we counted non-targeted Clarity flags. We distinguish these counterpart flags from \emph{co-flags}, which additionally require the intended C/A target itself to be flagged. A co-flag contributes one assessment to the non-targeted flag count. Conditions targeting both C and A do not enter this comparison, and cases with neither dimension targeted are described separately. This analysis identifies a pattern of joint ratings; it does not establish that one deficiency caused the other, particularly in conditions with additional intended perturbations.

Execution usage was measured as input plus output tokens accumulated across requests. For each of the eleven non-baseline executions, we calculate the percentage change in total tokens relative to the single baseline and summarize the unweighted mean of those changes. This measure describes accumulated processed token volume, including repeated cached input. 

Task outputs received an automated, unadjudicated checkpoint score. Seven binary checks assessed two group sizes, numerical activity estimates, comparison direction, the smallest gap, the number of significant comparisons, and absence of fabricated artifacts. Their weights were 2, 2, 3, 1, 1, 1, and 1, respectively, yielding a maximum of 11 points. We normalize the checkpoint total by dividing by 11 and multiplying by 100. The free-text checking procedure was not manually adjudicated. Scores therefore represent agreement with the automated checklist, whose treatment of equivalent ratio orientations, denominator definitions, and statistical choices requires separate verification.

\paragraph{Scope of inference.}
Scenario comparisons are descriptive: each condition has one execution of the same underlying task, and the five evaluator configurations assess those records. Additional assessments do not add task executions. All non-baseline comparisons share the same baseline run. Counts aggregated across evaluators describe repeated assessments of the same conditions, rather than independent experimental replications. We report condition-level outcomes, ordinal ratings, and threshold comparisons so that the aggregate results can be interpreted alongside the individual observations.

\section{Results}
\label{sec:results}

We report the three empirical components in the order presented in Methods. The survey analysis addresses the association between context problems and perceived AI usefulness. The naturalistic session evaluations characterize AX's measurement behavior. The scenario study examines resource use, automated task scores, and the correspondence between AX ratings and the intended context interventions.

\subsection{Context Problems and Perceived AI Usefulness}
\label{sec:results-so}

The initial survey eligibility rule retained 9,753 respondents. Of these, 336 met the reference-group definition and 7,847 reported at least one of the three specified context problems; respondents meeting neither definition were excluded from the between-group comparisons. Activity-specific denominators additionally excluded missing and ``Not applicable'' responses, as described in Section~\ref{sec:methods-so}.

\begin{figure*}[t]
  \centering
  \begingroup
  \definecolor{AXContextProblem}{HTML}{9DB9D6}
  \definecolor{AXContextReference}{HTML}{233C5B}
  \definecolor{AXContextText}{HTML}{20344E}
  \begin{tikzpicture}
    \begin{axis}[
      xbar,
      scale only axis,
      width=0.76\textwidth,
      height=0.42\textwidth,
      font=\small,
      xmin=0, xmax=52,
      ymin=-0.6, ymax=6.6,
      y dir=reverse,
      xtick={0,10,20,30,40,50},
      ytick={0,1,2,3,4,5,6},
      yticklabels={
        {Communication\,*},
        {Design\,*},
        {Implementation\,*},
        {Review\,*},
        {Debugging/refactoring\,*},
        {Architecture exploration},
        {Documentation\,*}
      },
      xlabel={Respondents rating AI ``Very useful'' (\%)},
      xlabel style={text=AXContextText},
      tick label style={text=AXContextText},
      ytick style={draw=none},
      axis lines=left,
      axis line style={-,draw=black!55,line width=0.35pt},
      tick style={draw=black!55},
      xmajorgrids=true,
      grid style={draw=black!12,line width=0.3pt},
      axis on top=false,
      bar width=7pt,
      point meta=x,
      nodes near coords={\pgfmathprintnumber[fixed,precision=1,zerofill]{\pgfplotspointmeta}\%},
      nodes near coords align=horizontal,
      every node near coord/.append style={
        anchor=west, xshift=2pt, font=\footnotesize,
        text=AXContextText
      },
      area legend,
      legend style={
        at={(0,1.03)}, anchor=south west,
        draw=none, fill=none, font=\small,
        legend columns=2,
        /tikz/every even column/.append style={column sep=1em}
      },
      legend cell align=left,
      clip=false
    ]
      \addplot+[
        fill=AXContextProblem,
        draw=AXContextReference!65,
        line width=0.3pt,
        bar shift=4pt
      ] coordinates {
        (14.1,0) (20.2,1) (35.3,2) (29.8,3)
        (35.8,4) (27.2,5) (39.0,6)
      };
      \addlegendentry{Reports context problems}

      \addplot+[
        fill=AXContextReference,
        draw=AXContextReference,
        line width=0.3pt,
        bar shift=-4pt
      ] coordinates {
        (19.6,0) (28.0,1) (44.4,2) (36.8,3)
        (42.8,4) (32.2,5) (44.5,6)
      };
      \addlegendentry{Reports adequate context (reference)}
    \end{axis}
  \end{tikzpicture}
  \endgroup
  \caption{Perceived AI usefulness by development activity and context group in the Stack Overflow Developer Survey~\cite{stackoverflow2026}. The reference group reported adequate context ($n=336$); the problem group reported outdated, incomplete, or conflicting information ($n=7{,}847$). Activity-specific denominators exclude missing and ``Not applicable'' responses. Percentages retain the precision of the recorded analysis. An asterisk denotes an unadjusted 95\% confidence interval for the ratio of the problem-group proportion to the reference-group proportion that excludes one. These observational associations do not establish causality.}
  \label{fig:so-context-results}
\end{figure*}
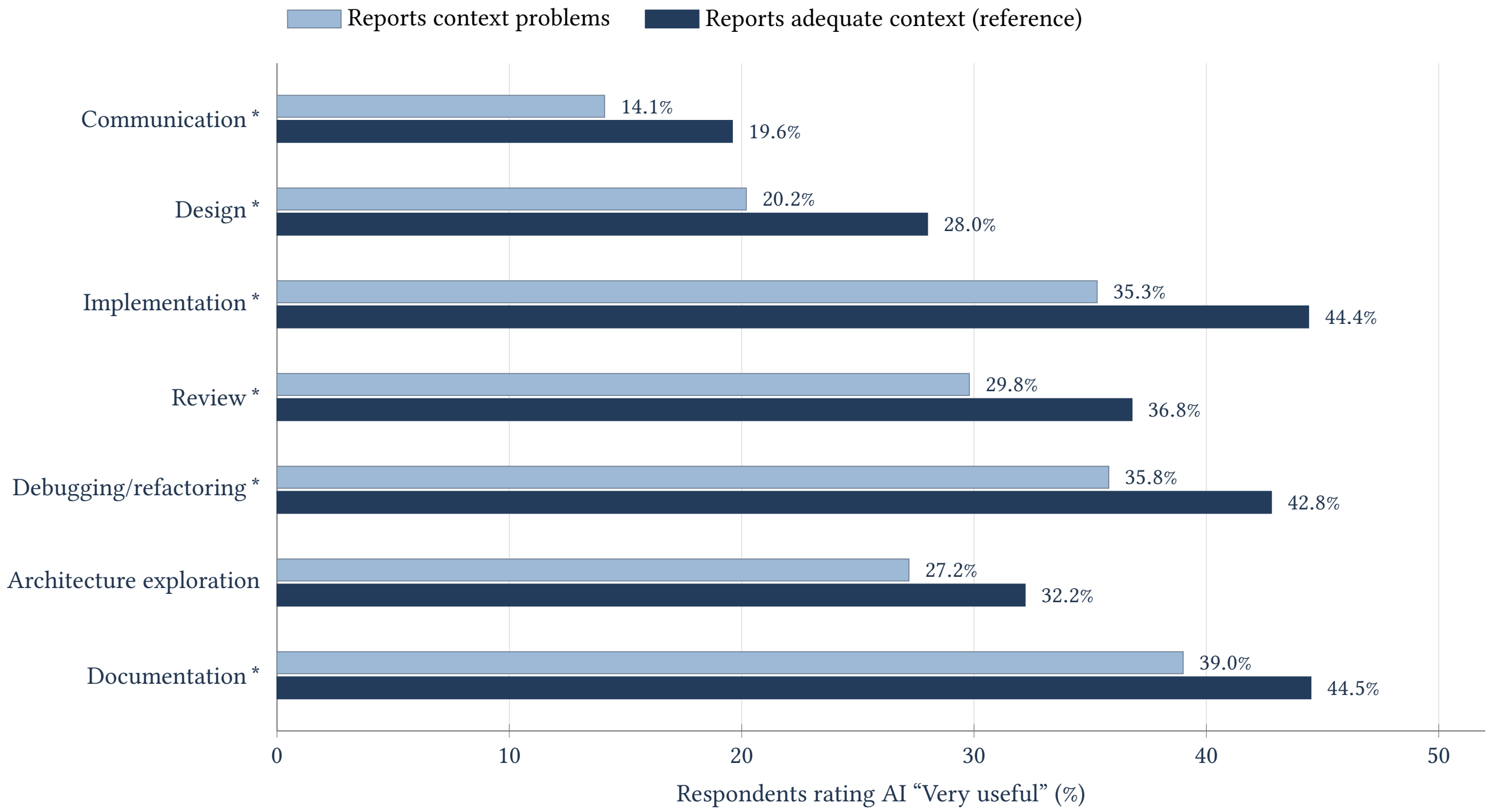

Across all seven activities, respondents reporting context problems were less likely to rate AI as ``Very useful'' (Figure~\ref{fig:so-context-results}). The ratio of the context-problem proportion to the reference proportion ranged from 0.718 for communication to 0.876 for documentation, corresponding to relative differences of 28.2\% and 12.4\%, respectively. Absolute differences calculated from the reported, rounded percentages ranged from 5.0 to 9.1 percentage points. Six of the seven comparisons were marked as having unadjusted 95\% confidence intervals excluding one in the primary analysis. Architecture exploration was the exception, although its point estimate was in the same direction.

% Native PGFPlots figure for all_files/ax_so_context_figure.tex.
% The updated Results section inputs this file in place of the SO table.
% Requires pgfplots with compat=1.18, already loaded in the supplied main.tex.
% All plotted values come from derived/survey_results_reported.csv.
% No external image or CSV is required to compile this figure.

For the observational component of RQ1, the direction of association was consistent across activities: reported information problems accompanied lower perceived AI usefulness. These comparisons concern respondents' reports of their working context and AI use. They do not identify the context supplied to a particular agent session or estimate the causal effect of changing that context.

\subsection{AX Measurement Behavior in Naturalistic Sessions}
\label{sec:results-naturalistic}

\paragraph{Coverage and use of the response scale.}
The study sampled 1,000 session transcripts from the DX Agent Experience pilot collection. The retained evaluation outputs contained 968 distinct session identifiers. The five-point baseline produced numerical ratings for 882 sessions on each dimension. The other 86 sessions did not contribute numerical ratings; the more detailed repeat-run record distinguished 73 sessions with insufficient evidence from 13 with nonnumeric dimension assessments.

The baseline ratings varied across both sessions and dimensions (Table~\ref{tab:ax-baseline-distributions}). Mean scores were 3.60 for Clarity, 3.34 for Actionability, 4.09 for Fidelity, and 3.74 for Efficiency. Between 26.6\% and 59.6\% of ratings, depending on the dimension, were at or below 3. Maximum ratings ranged from 6.3\% for Actionability to 39.1\% for Fidelity. Clarity and Actionability used all five response categories; Fidelity and Efficiency used categories 2--5. The baseline therefore showed substantial variation below the upper two categories, with the greatest concentration at the maximum in Fidelity.

\begin{table}[t]
  \centering
  \small
  \caption{Five-point baseline AX ratings in naturalistic sessions. Each dimension has 882 numerical ratings. Means and percentages were recalculated from the recorded category counts. Higher scores indicate less context-related friction.}
  \label{tab:ax-baseline-distributions}
  \begin{tabularx}{\columnwidth}{@{}Xrrr@{}}
    \toprule
    Dimension & Mean & \shortstack{Score $\leq 3$\\(\%)} & \shortstack{Score 5\\(\%)} \\
    \midrule
    Clarity & 3.60 & 39.6 & 8.2 \\
    Actionability & 3.34 & 59.6 & 6.3 \\
    Fidelity & 4.09 & 26.6 & 39.1 \\
    Efficiency & 3.74 & 35.6 & 16.9 \\
    \bottomrule
  \end{tabularx}
\end{table}

\paragraph{Relationships among ratings.}
The reported pairwise dimension correlations ranged from 0.404 to 0.678. Clarity and Actionability had the largest coefficient (0.678), followed by Actionability and Efficiency (0.646); Clarity and Fidelity had the smallest (0.404). The dimensions thus produced associated but nonidentical ratings.

Ratings were also negatively associated with session message count. The reported Spearman coefficients were $-0.213$, $-0.344$, $-0.396$, and $-0.529$ for Clarity, Actionability, Fidelity, and Efficiency, respectively. Efficiency had the strongest association with session length. This pattern is consistent with greater navigation and recovery effort in longer sessions, although task complexity could also contribute.

\paragraph{Recorded agreement and question perspective.}
Runs 1--3 produced identical numerical scores and explanatory comments, differing only in how insufficient evidence and input warnings were represented. Section~\ref{sec:limitations} discusses what this agreement does and does not show about repeatability.

Changing the question perspective from second person to first person produced similar aggregate means: all four reported means differed by less than 0.05 points. Session-level ratings nevertheless changed. Exact agreement ranged from 63.7\% for Fidelity to 78.8\% for Actionability, and agreement within one point ranged from 96.9\% to 100\%. These comparisons used 880 matched numerical ratings for Fidelity and 881 for each other dimension. Similar aggregate averages therefore coexisted with variation in individual assessments.

\paragraph{Response scale, anchors, and item count.}
The anchored three-point configuration and the first-person five-point configuration had reported rank correlations of 0.48--0.55 across dimensions. On the three-point scale, 712 of 896 Actionability ratings (79.5\%) occupied the middle category, compared with 127 (14.2\%) in the highest category. The three-point rubric concentrated many assessments in its middle response option.

The comparison with the unanchored three-point configuration reported identical ratings across all four dimensions for 468 of 894 jointly scored sessions (52.3\%). Dimension-level exact agreement ranged from 81.9\% to 86.6\%. The reported matched-comparison means changed most for Efficiency, from 2.65 to 2.55.

The two-question, five-point configuration produced mean dimension scores of 3.34, 3.10, 3.77, and 3.83 for Clarity, Actionability, Fidelity, and Efficiency, respectively. Clarity, Actionability, and Efficiency each had 898 numerical dimension scores; Fidelity had 887. Rank correlations with the first-person, single-question configuration ranged from 0.63 to 0.73. The reported means for Clarity, Actionability, and Fidelity were lower under the two-question configuration, while Efficiency was slightly higher. The configurations produced different absolute scores even when session rankings retained a positive association.

\paragraph{Evaluator configuration.}
The five-point evaluations also differed across evaluator configurations. The first-person Auto configuration had reported means of 3.57, 3.30, 4.07, and 3.75 across CAFE. The corresponding means were 4.06, 3.65, 3.05, and 3.29 for GPT-6 Luna, and 3.54, 3.11, 4.07, and 3.69 for GPT-6 Sol. Fidelity showed the largest descriptive difference between Luna and the Auto configuration, approximately one scale point. Because the Luna run also changed transcript handling and instructions, part of this difference may reflect those changes rather than the model.

\paragraph{Manual review of baseline output.} All 15 sessions pass on all three judgments. Because the reviewer was a single individual who saw AX's output alongside the transcripts, this result is a preliminary plausibility check and not an accuracy test. 

For RQ2, AX produced differentiated ratings that used the response scale and separated the dimensions. Aggregate scores were stable when the question perspective changed (all four means within 0.05 points), while response scale, item formulation, and evaluator configuration shifted absolute scores in some dimensions. A manual review of 15 sessions by one researcher found no issues with the scores, explanations, or suggested improvements. In practice, AX scores should be compared within a fixed configuration.

\subsection{Scenario Execution Measures}
\label{sec:results-execution}

\paragraph{Processed token volume.}
The clean baseline used 473,829 total input and output tokens. Across the eleven intended defect conditions, mean usage was 905,534 tokens, an increase of 91.1\% relative to that baseline. Every non-baseline execution used more total tokens, with increases ranging from 39.7\% in AF to 196.3\% in CE (Table~\ref{tab:scenario-execution-results}). The median non-baseline execution used 835,238 tokens, 76.3\% above the baseline.

\begin{table}[t]
  \centering
  \small
  \caption{Scenario execution measures. Condition letters denote intended CAFE interventions. Each condition has one execution of the same underlying task. Token changes share the single CLEAN baseline. The checkpoint score is an automated, unadjudicated measure with a maximum of 11.}
  \label{tab:scenario-execution-results}
  \begin{tabularx}{\columnwidth}{@{}Xrrr@{}}
    \toprule
    Condition & \shortstack{Total\\tokens} & \shortstack{Change\\(\%)} & \shortstack{Checkpoint\\score /11} \\
    \midrule
    CLEAN & 473,829 & --- & 9 \\
    C & 1,075,279 & $+126.9$ & 5 \\
    A & 835,238 & $+76.3$ & 5 \\
    F & 673,934 & $+42.2$ & 5 \\
    E & 714,082 & $+50.7$ & 5 \\
    CA & 896,134 & $+89.1$ & 5 \\
    CF & 827,761 & $+74.7$ & 5 \\
    CE & 1,403,818 & $+196.3$ & 10 \\
    AF & 662,066 & $+39.7$ & 5 \\
    AE & 676,985 & $+42.9$ & 5 \\
    FE & 944,727 & $+99.4$ & 6 \\
    ALL & 1,250,852 & $+164.0$ & 6 \\
    \midrule
    Non-baseline mean & 905,534 & $+91.1$ & 5.64 \\
    \bottomrule
  \end{tabularx}
\end{table}

Input usage dominated these totals and included substantial repeated cache reads. The increase measures accumulated processed token volume; billed cost and elapsed time were not estimated. Output volume alone would give a different picture in at least one condition: E produced 4,002 output tokens, fewer than the baseline's 5,078, despite using 50.7\% more total tokens.

\paragraph{Automated checkpoint scores.}
The baseline received 9 of 11 checkpoint points (81.8\%). The other executions averaged 5.64 of 11 points (51.2\%), a difference of 30.6 percentage points. Eight conditions received 5 points, FE and ALL received 6, and CE received 10. CE therefore had both the highest checkpoint score and the greatest token usage.

The individual checks qualify the meaning of the aggregate difference. All twelve executions passed the numerical activity-estimate check, while none passed the check requiring exactly six significant comparisons. The execution audit identified possible false failures caused by reciprocal ratio orientation, differences in reported group denominators, and alternative significance procedures. The 30.6-point gap therefore measures agreement with the automated checklist, not task accuracy.

For the scenario component of RQ1, every defect condition used more tokens than the baseline, while all twelve executions produced correct activity-level estimates. In this task, context defects appeared as added cost before they appeared as error.

\subsection{AX Ratings and Intended Context Interventions}
\label{sec:results-diagnostic}

Five evaluator configurations (Opus~5.5, Haiku~4.5, Haiku~5.5, Luna~6, and Terra~6.1) each supplied 48 dimension ratings for the twelve scenario transcripts. Under the intended condition matrix, 20 ratings in each series concerned targeted dimensions and 28 concerned non-targeted dimensions, including CLEAN's four ratings. We recomputed diagnostic flags using Equation~\ref{eq:ax-scenario-flag}: a score below 3 or a decline of at least two categories from the same evaluator's CLEAN score on that dimension.

\begin{table}[t]
  \centering
  \small
  \caption{Scenario diagnostic flags: score below 3 or at least two categories below the corresponding CLEAN score. The non-targeted denominator includes CLEAN. Labels describe intended interventions.}
  \label{tab:scenario-ax-flags}
  \begin{tabularx}{\columnwidth}{@{}Xrr@{}}
    \toprule
    Evaluator & \shortstack{Targeted flags\\$n/20$ (\%)} & \shortstack{Non-targeted flags\\$n/28$ (\%)} \\
    \midrule
    Haiku 4.5 & 10/20 (50.0) & 9/28 (32.1) \\
    Haiku 5.5 & 12/20 (60.0) & 4/28 (14.3) \\
    Opus 5.5 & 14/20 (70.0) & 6/28 (21.4) \\
    Luna 6 & 17/20 (85.0) & 9/28 (32.1) \\
    Terra 6.1 & 17/20 (85.0) & 10/28 (35.7) \\
    \bottomrule
  \end{tabularx}
\end{table}

\paragraph{Correspondence with targeted dimensions.}
The five evaluators flagged 50.0--85.0\% of intended targeted assessments (Table~\ref{tab:scenario-ax-flags}). Haiku~4.5 flagged 10 of 20, Haiku~5.5 flagged 12, Opus~5.5 flagged 14, and Luna and Terra each flagged 17.

The ordinal profiles provide additional context. For each evaluator, the pooled mean rating was lower for targeted than for non-targeted assessments, with differences of 0.44--1.41 scale points. Fidelity had the largest targeted versus non-targeted mean separation within every evaluator (1.09--2.57 points). Actionability had the weakest separation for every evaluator except Opus~5.5, for which Efficiency separated least (0.34 points). Targeted Actionability scores were lower by 0.26--0.74 points for four evaluators, while Haiku~4.5 scored targeted Actionability assessments 0.31 points higher on average.

\paragraph{Non-targeted flags and baseline ratings.}
Non-targeted flag rates were 14.3--35.7\% across the five evaluators (21.4\% for Opus~5.5). All CLEAN ratings were 4 or 5, so none met the diagnostic criterion. The CLEAN profiles, ordered C/A/F/E, were 5/5/5/5 for Luna, 5/5/4/5 for Opus~5.5, 5/5/5/4 for both Haiku configurations, and 4/4/5/5 for Terra. In these records, every score below 3 was also at least two categories below CLEAN; the combined rule therefore reduced to a two-category decline. Relative to the absolute below-3 rule, it additionally flagged scores of 3 where CLEAN was 5.

The distinction between intended and delivered interventions is material to these counts. AE was intended to target Actionability and Efficiency, but its delivered prompt included the false factual statements and omitted the intended efficiency material. Haiku~4.5, Opus~5.5, Luna, and Terra each assigned Fidelity a 2 in this session. Those scores count as non-targeted flags under the intended matrix, although they correspond to a deficiency documented in the delivered prompt. Non-targeted flags can include additional deficiencies present in the session as well as potentially unwarranted deductions; the present records do not independently classify all such cases.

\paragraph{Clarity and Actionability co-flags.}
Across the five evaluators, 13 of the 38 non-targeted flags occurred on one C/A dimension when the other was an intended target. Twelve of these also accompanied a flag on that intended counterpart: these co-flags accounted for 12/38 (31.6\%) of all non-targeted flags. The exception was Haiku~4.5 in CF, where Actionability scored 3 and was flagged, while targeted Clarity scored 4 and was not. The co-flag counts were 1/9 for Haiku~4.5, 3/4 for Haiku~5.5, 2/6 for Opus~5.5, 3/9 for Luna, and 3/10 for Terra, with denominators representing each evaluator's non-targeted flags.

The pattern was asymmetric. In conditions targeting Clarity without Actionability (C, CF, CE), the five evaluators flagged non-targeted Actionability in 11 of 15 assessments, with both dimensions flagged in ten. In conditions targeting Actionability without Clarity (A, AF, AE), non-targeted Clarity was flagged in two of 15 assessments, both accompanied by a flag on the intended target. Haiku~4.5 also flagged both C and A in FE, where neither was targeted; that pair contributes two additional non-targeted flags and is kept separate from the one-target co-flag analysis.

\paragraph{Threshold comparisons.}
The operational definition affected both targeted and non-targeted flagging. Among the five evaluators, absolute scores below 3 yielded targeted rates of 20.0--80.0\% and non-targeted rates of 0.0--21.4\%. Any decline from CLEAN yielded 75.0--100.0\% and 42.9--60.7\%, respectively. Scores below 5 yielded 80.0--100.0\% and 50.0--78.6\%, respectively, including the baseline's nonmaximum scores. All comparisons retain denominators of 20 and 28. The combined criterion thus captures a different degree of rating change from either ordinary friction or absolute high friction alone.

For RQ3, AX ratings tracked the intended interventions, most consistently for Fidelity and least for Actionability. Evaluator choice moved targeted detection from 50\% to 85\%. Clarity--Actionability co-flags were the most common non-targeted response. Section~\ref{sec:limitations} discusses how label visibility and imperfect isolation bound these rates.

\section{Limitations}
\label{sec:limitations}

This study provides the first empirical account of AX as an operationalization of CAFE. Its three components address different questions: the survey examines associations with perceived AI usefulness, the naturalistic evaluations characterize rating behavior, and the scenario study examines responses to deliberate context perturbations. The limitations below define the scope of each.

\subsection{Internal Validity}

\paragraph{Researcher involvement and scenario execution.}
An author involved in developing CAFE(S) designed the scenario interventions and enacted the human participant while knowing the expected analysis and evaluation criteria. This involvement introduces the possibility of researcher expectancy influencing scenario design, follow-up dialogue, and interpretation. Authored initial conditions and an interaction policy provided structure, but subsequent dialogue was adaptive. The execution audit also identified departures from the planned interventions, including changes to the delivered deficiencies and later provision of initially missing information. The comparisons therefore capture the resulting interaction and recovery trajectories, with incomplete isolation of the initial context perturbations. We document these departures in Section~\ref{sec:scenario-methods}.

Corrective documentation and prior analysis artifacts were accessible in the shared workspace, and the README was accessed in nine of the twelve executions, including the baseline. Access to this material could support recovery from introduced problems while also changing search and verification effort. Resource and outcome comparisons therefore reflect sessions in which this material was available. Descriptive acronyms were visible in the condition names to four of the five evaluators, and human clarifications could cue any evaluator to intended deficiencies. The scenario results therefore do not isolate blinded diagnostic performance.

\paragraph{Observational associations.}
The survey analysis compares self-reported context problems with self-reported AI usefulness. It is cross-sectional and does not adjust for respondent, task, or organizational characteristics that could affect both measures. Differences in experience, work requirements, or organizational support could contribute to the observed associations. The survey findings support the relevance of the context hypothesis without identifying a causal effect of context quality.

\subsection{Construct Validity}

\paragraph{Evidence available to evaluators.}
The naturalistic evaluations used exported user--assistant conversations. The executing agents' original system instructions, structured tool interactions, and referenced materials were available to evaluators only when reproduced in the messages. Evaluators could therefore assess conversational evidence of ambiguity, clarification, correction, and recovery, while some conditions encountered during execution remained unobservable. Absence from the export does not establish absence during the original interaction. This boundary particularly affects independent verification of Fidelity and the distinction between inaccurate supplied context and errors introduced by the executing agent. Accuracy of scores was assessed only through a manual review of 15 randomly sampled sessions by a single researcher who saw AX's score and explanation before review. This design is susceptible to confirmation bias and provides no inter-rater reliability. Independent and blinded reviews of a larger sample size is still needed.

\paragraph{Dimensions and diagnostic labels.}
CAFE supplies the instrument's conceptual structure, but differentiated score distributions and correlations do not by themselves establish discriminant validity or a four-factor measurement structure. The dimensions can also co-occur within a session. Scenario labels describe intended interventions, and execution departures mean that they are not verified diagnoses of every dimension in the resulting records. Consequently, a non-targeted flag can reflect an additional deficiency, an effect spanning dimensions, or an unwarranted deduction. The observed Clarity--Actionability co-flagging is compatible with overlapping consequences of an ambiguous request, but does not establish that one deficiency caused another or that all additional flags were warranted. Mixed interventions and condition labels visible to most evaluators provide alternative explanations. We therefore report targeted and non-targeted flags as descriptive correspondence measures, with precise localization remaining a question for independent reference assessment.

\paragraph{Execution measures.}
The automated checkpoint score was not manually adjudicated and can reject semantically equivalent answers or alternative analytical choices. Its differences represent agreement with the implemented checklist, rather than validated differences in task accuracy. Total tokens measure accumulated input and output volume, including repeated cached input. They do not directly measure billed cost, elapsed time, or productivity.

\paragraph{Scope of assessment.}
AX characterizes evidence of CAFE-related friction in the available record. Favorable ratings do not establish that the underlying human intent was appropriate or complete, nor do they provide security assurance.

\subsection{External Validity}

The naturalistic corpus was randomly sampled from the available DX Agent Experience pilot collection. Random selection within that collection does not establish that pilot participants, supported tools, or their task distributions represent agent use more broadly. Inclusion in the collection and the availability of sufficient evidence for numerical scoring also constrain the analyzed population. The resulting score distributions describe the evaluated corpus; they should not be treated as estimates of the prevalence of CAFE deficiencies across developers or organizations generally.

The scenario study used one data-analysis task, one executor configuration, and one researcher acting as the human participant. Its twelve conditions represent variations of that task. The authored perturbations were deliberately explicit, providing salient cases for examining the instrument's response. Performance on subtler, gradually emerging, or otherwise different deficiencies may differ. The five evaluator configurations broaden the observed range of assessment behavior, while generalization to other tasks, interaction styles, and deployment environments remains to be established. The evaluated instrument covers Clarity, Actionability, Fidelity, and Efficiency; Security was outside this study's scope.

The study did not evaluate whether presenting AX feedback changes human behavior, improves subsequent context, or produces better downstream outcomes. The feedback loop proposed in the Discussion is therefore a prospective application requiring longitudinal evaluation.

\subsection{Conclusion Validity}

Each scenario condition was executed once, preventing estimation of within-condition stochastic variability. All resource-use contrasts share a single baseline execution, and all five scoring series assess the same twelve records. Their ratings provide alternative assessments of those executions, rather than additional independent task replications. Each evaluator's score comparisons likewise share one retained CLEAN rating per dimension. Scenario means, score contrasts, and flag rates are therefore descriptive summaries of the recorded conditions. The magnitude of the token increase should not be interpreted as a population effect estimate. The survey confidence intervals were not adjusted for the seven comparisons. The activities also do not constitute independent replications because respondents could contribute to multiple outcomes.

The diagnostic flag criterion was selected during exploratory analysis to recognize either high friction or a decline of at least two categories from CLEAN. Such differences are contrasts between ordinal rubric categories, not established equal-interval changes or a validated minimum reliable difference. Alternative-criterion comparisons make threshold dependence visible without independently validating the selected cutoff. CLEAN is included in the non-targeted denominator of 28 assessments per evaluator and also supplies the reference scores. Its comparison with itself cannot trigger the relative-decline branch, and all CLEAN ratings were 4 or 5, above the absolute high-friction threshold. The resulting rate describes non-targeted flagging across the full set of runs, including that reference condition.

For the naturalistic evaluations, exact agreement among the retained baseline outputs does not establish independent-run repeatability because the provenance of fresh model calls was not established. Some evaluator comparisons also changed transcript handling or evaluation instructions, and numerical eligibility varied across configurations. Observed differences therefore characterize the implemented configurations; they cannot always be attributed solely to the evaluator model or a single prompt feature. We therefore report rating behavior per configuration and treat independent repeatability and cross-configuration calibration as next steps (Sections~\ref{sec:fw-validation} and~\ref{sec:fw-calibration}).

\section{Discussion}
\label{sec:discussion}

\subsection{From Context Properties to Session Assessment}

CAFE(S) provides a vocabulary for the properties of useful context~\cite{houck2026cafes}. AX makes the first four properties operational through explicit questions, scoring criteria, and explanations grounded in recorded interactions. Three kinds of evidence support this procedure. Developers who reported context problems found AI less useful in every activity examined; in naturalistic sessions, AX ratings used the response scale and separated the dimensions, and a manual review of 15 sampled sessions found no issues; and in the scenario study, ratings tracked deliberately introduced context deficiencies. Together, these components demonstrate the feasibility of AX as a concrete approach to measuring evidence of CAFE-related friction during agentic work.

The scenario findings show responsiveness to context perturbations while distinguishing it from precise localization. Under the exploratory criterion of a score below 3 or a decline of at least two rubric categories from CLEAN, the five evaluators flagged 50--85\% of intended targeted dimensions and 14.3--35.7\% of non-targeted dimensions (Section~\ref{sec:results-diagnostic}). Because the interventions did not perfectly isolate the dimensions, non-targeted flags can represent additional deficiencies or effects spanning properties, as well as unwarranted deductions. The dimension profile and supporting explanation are therefore necessary for interpreting a flag and deciding what to investigate.

The naturalistic and scenario results point to the same priority. Actionability received the lowers scores in pilot sessions (59.6\% of ratings at or below 3) and the dimension evaluators detected least consistently in the scenario (13 of 25 targeted assessments flagged across the five evaluators), while Fidelity received the highest scores and was the most reliably detected (22 of 25). Part of the scenario gap reflects how the Actionability interventions were delivered, since completion criteria were supplied later in the session in each Actionability condition. Together, these results identify Actionability, whether the human states an objective, constraints, and a definition of done, as both the largest opportunity for teams and the most important target for improving the instrument.

The criterion captures two reasons an assessment can be useful: high friction in absolute terms, or deterioration relative to the clean reference. In particular, a change from 5 to 3 moves from effortless progress to real but bounded recovery, even though 3 is the rubric's baseline category. Detecting such a change can reveal additional work imposed by the information environment before the interaction becomes severely impaired. Because the cutoff was chosen during exploratory analysis (Section~\ref{sec:limitations}), Section~\ref{sec:results-diagnostic} also reports results under alternative criteria.

\subsection{Interpreting Clarity and Actionability Together}

Clarity and Actionability account for a recognizable subset of the additional flags (Section~\ref{sec:results-diagnostic}). Across the five evaluators, 13 of 38 non-targeted flags occurred in the other member of this pair when exactly one was an intended target. In twelve of those instances, the intended target was also flagged: these paired responses accounted for 31.6\% of non-targeted flags. The pattern was asymmetric: Actionability was flagged in 11 of 15 assessments where Clarity was targeted and Actionability was not, whereas Clarity was flagged in two of 15 assessments with the reverse targeting. Each direction comprises repeated assessments of the same three conditions.

This pattern is compatible with the conceptual relationship between the properties. An ambiguous term can affect interpretation of a request and the ability to identify an executable objective or completion criterion. The rubric accordingly permits one event to support deductions in more than one dimension, provided each deduction has evidence. The useful diagnostic question is whether the record supports both assessments. Clarity and Actionability need not be independent in every interaction to remain separately interpretable properties.

Section~\ref{sec:limitations} discusses alternative explanations for the pairing, including mixed interventions and visible condition labels. In practice, the pattern identifies a calibration target: rubric guidance on when a single event warrants deductions in both dimensions (Section~\ref{sec:fw-calibration}). It does not support collapsing Clarity and Actionability into one construct.

\subsection{AX Within a Broader Measurement Practice}

AX addresses one part of understanding agentic effectiveness. Teams also need evidence about the availability of information, the capabilities of their configured systems, and the quality and resource demands of the resulting work. Table~\ref{tab:complementary-measurements} distinguishes these questions as a proposed framework; this study evaluates the AX row.

\begin{table*}[t]
\centering
\caption{Complementary questions for measuring agentic work. AX contributes session-level evidence within this broader practice.}
\label{tab:complementary-measurements}
\small
\begin{tabularx}{\textwidth}{@{}p{0.20\textwidth}p{0.31\textwidth}X@{}}
\toprule
Measurement focus & Question & Illustrative evidence \\
\midrule
Context and execution preconditions & Does the necessary information exist, and can the system obtain and use it? & Explicit objectives and acceptance criteria; maintained source material; resolvable references; required tool access. \\
\addlinespace
Model, harness, and tool evaluations & Can the configured system perform representative tasks under expected operating conditions? & Model and tool tests; retrieval evaluations; end-to-end task evaluations with recorded configurations. \\
\addlinespace
AX session assessment & What evidence of context-related friction appears during the interaction? & CAFE ratings and supporting explanations; clarification, correction, search, and recovery visible in the session record. \\
\addlinespace
Outcomes and resource use & Did the work meet its requirements, and what did execution require? & Independently checked results; review and rework; elapsed time; billed cost and processed tokens; subsequent maintenance where relevant. \\
\bottomrule
\end{tabularx}
\end{table*}

The scenario executions illustrate why these questions should be examined together. All twelve executions passed the numerical activity-estimate checkpoint, while all eleven non-baseline executions processed more tokens than the baseline (Section~\ref{sec:results-execution}). A core answer check therefore coexisted with substantial differences in processed token volume. Processed token volume rose by 91.1\% on average, including cached input; Section~\ref{sec:limitations} distinguishes this from billed cost. AX contributes evidence about the interaction through which an answer was reached, while independent outcome and resource measures establish other aspects of that execution.

This division of responsibilities also places AX alongside evaluation of the harness, tools, and model. Poor retrieval, unavailable tools, or limited model capability can affect what appears in a session, and an AX score alone cannot attribute a problem to one component. Conversely, a capable system can receive inadequate instructions or misleading source material. AX ratings describe evidence from a task under a particular configuration and observation boundary. They should be interpreted with information about that configuration and the context actually made available.

\subsection{Context Quality Depends on the Intent It Represents}

CAFE(S) situates context between human intent and the harness and model that act on it, and distinguishes assembled context from prerequisites such as knowledge capture and retrieval~\cite{houck2026cafes}. This distinction matters for interpreting AX. An evaluator may identify an ambiguous objective, an unstated constraint, or missing completion criteria when these deficiencies are visible in the record. It cannot reliably reconstruct requirements that were never expressed or determine, from a transcript alone, whether the requested objective was the right organizational choice.

A clear, actionable request can still pursue an unsuitable goal. Favorable AX ratings therefore do not establish that the underlying intent is complete, valuable, or appropriate. Teams must continue to elicit requirements, resolve stakeholder tradeoffs, and check whether the resulting work serves the intended need. Clarifying a missing requirement can itself be useful work. A measurement practice should distinguish necessary clarification from avoidable friction, rather than treating every question or additional interaction as evidence of poor performance.

Precondition checks can contribute to this interpretation. For example, a deterministic check may establish that a required field is present or that a referenced artifact exists. Those checks do not establish whether the field expresses the right objective or whether the artifact is sufficient for the task. Combining such checks with contextual assessment is a promising extension, developed in Section~\ref{sec:future-work}.

\subsection{A Feedback Loop for Human Learning and Better Context}

AX could support a feedback loop in which evidence from today's sessions informs the context used in subsequent work. An assessment might draw attention to an ambiguous term, a repeatedly omitted constraint, or a conflicting reference. A developer could review that evidence, correct the source of the problem, and preserve the correction in a specification, glossary, example, or other shared artifact. Later sessions would then provide evidence about whether the change addressed the recurring difficulty.

This proposed use connects to the two-way scaffolding described by Choudhuri et al.~\cite{choudhuri2026boldly}. Their account emphasizes developers learning with AI while externalizing knowledge and standards that guide AI's work. AX could help make the need for that externalization visible. The guidelines for responsible AI integration likewise connect clarification of missing context, maintenance of shared knowledge, and independent verification of AI contributions~\cite{choudhuri2026guidelines}. These connections motivate a role for AX in supporting human judgment and context maintenance.

The feedback loop requires interpretation and follow-through. A low score becomes useful when someone can inspect the evidence and make an appropriate change. Automatically copying an evaluator's explanation into future context could instead preserve an incorrect inference or introduce unnecessary material. The proposed loop therefore involves reviewed changes to context, followed by assessment of subsequent work against independent outcomes. It does not require training the underlying model. Whether this loop improves human understanding, later context, and task outcomes is the subject of Section~\ref{sec:fw-feedback}.

\subsection{Why Security Requires a Distinct Assessment Approach}

The exclusion of Security is a deliberate boundary of the present instrument. CAFE(S) treats Security as a distinct question about whether context is safe and appropriate to use~\cite{houck2026cafes}. A session can be fluent and apparently successful while handling information that should not have been available or following instructions from an untrusted source. Conversely, a blocked action or additional verification may indicate that a protective control worked. A scale designed to characterize context-related friction cannot resolve those distinctions without security-specific criteria and evidence.

The consequences of a missed security failure can also demand stronger assurance than is appropriate for routine diagnostic feedback. Depending on the task, such a failure could expose sensitive information or permit an unauthorized action. Extending AX to Security therefore requires an explicit threat model, suitable observations, and validation that accounts for the consequences of false negatives. Favorable CAFE ratings do not provide security assurance, and strengths in the other dimensions should not compensate for an identified security failure in an overall score. Security is a separate research direction within the broader goal of measuring context health.

\subsection{Using Scores to Support Improvement}

The configuration differences observed in the naturalistic evaluations argue for recording the evaluator model, prompt version, and available evidence when interpreting AX scores. Comparisons across configurations require calibration; missing evidence should remain distinguishable from favorable evidence. Session ratings also depend on task demands and system behavior: AX is designed to diagnose context, not to rank individual developers.

The useful target is better work supported by better context. Improvements in AX scores should therefore be assessed alongside independently checked outcomes and resource use. A session with fewer visible corrections is not necessarily better if the agent stopped asking necessary questions or the record omitted the relevant evidence. Maintaining that distinction allows AX to support diagnosis and learning without making the score itself the objective.

% Complete Future Work section for all_files/7.Future_Work.tex.
% Uses the existing Methods/Results labels and the Discussion label.
% These are prospective studies, not additional work claimed in this paper.

\section{Future Work}
\label{sec:future-work}

The present study establishes an implemented AX instrument and an initial account of its behavior. The next steps are to strengthen the interpretation of its ratings and evaluate how those ratings can improve human--agent work.

\subsection{Diagnostic Validation Across Tasks and Settings}
\label{sec:fw-validation}

A priority is to develop reference assessments of the deficiencies actually present in session records. Intended scenario labels are insufficient when interventions overlap, additional deficiencies emerge, or later dialogue repairs an initial problem. Independent assessors should evaluate each CAFE dimension using documented criteria, with uncertainty and disagreement retained or adjudicated. Evaluator models should be blinded to condition labels. These reference assessments would support evaluating missed deficiencies and unwarranted flags at explicit thresholds, while distinguishing incorrect assessments from real deficiencies outside the intended intervention.

The reference set should include verified absence of a deficiency, subtle and severe cases, productive clarification, recovery, and records with insufficient evidence. Broader studies should vary tasks, human participants, executor configurations, and naturally occurring context problems. Repeated executions and independently logged evaluator calls, including repeated CLEAN conditions, would allow estimation of execution variability, assessment repeatability, and the stability of baseline-relative differences. These extensions would also clarify whether observed score differences reflect distinct CAFE properties, common task difficulty, or the evidence available to the evaluator.

The Clarity--Actionability pattern motivates a focused comparison of isolated and coupled deficiencies. For example, tasks could vary the ambiguity of necessary terms separately from the completeness of objectives and acceptance criteria, then include conditions in which both are deficient. Reference assessors would document whether each property was affected in the delivered interaction. This would help distinguish warranted co-flagging from a tendency to carry one deduction across dimensions, while preserving the conceptual distinction between understanding a request and being able to act on it. Parallel comparisons could assess blinded and label-visible scoring of the same transcripts under matched evidence.

\subsection{Prompt Refinement and Calibration From Field Feedback}
\label{sec:fw-calibration}

Real-world use can inform refinement of the questions, anchors, and requested explanations. Feedback should include cases where a rating was unhelpful, the explanation lacked support, or an assessment confused an agent-generated error with a deficiency in supplied context. The current scenario results suggest particular value in clarifying when Clarity and Actionability should receive separate deductions and when the same event provides evidence for both. Refinement should preserve legitimate overlap without collapsing the dimensions or requiring them to vary independently in every task.

Prompt development should use a versioned development set and a separate evaluation set that is not used to revise the instrument. The diagnostic criterion and denominator rules should be fixed before evaluating new held-out records. The present rule, identifying high friction or a decline of at least two rubric categories, is a candidate for prospective calibration. Comparisons should assess diagnostic correspondence, severity calibration, evidence quality, and appropriate abstention, alongside agreement. An insufficient-evidence response should not be collapsed into a score indicating healthy context. Changes to the prompt or evaluator should be accompanied by assessment on a common reference set so that a shift in the measurement procedure is not mistaken for a change in context quality. This agenda concerns prompt and rubric refinement; model-weight fine-tuning is a separate possibility that the present study does not evaluate.

\subsection{Combining Deterministic Checks With AX}

A complementary direction is to combine AX with deterministic checks of context artifacts and execution preconditions. Candidate checks include required fields, resolvable references, schema conformance, explicit artifact-version mismatches, and violations of specified size limits. Such checks could identify some problems before execution and supply additional evidence when interpreting a session afterward.

Their scope must remain explicit. The existence of a document does not establish that it contains the needed knowledge, and a recent modification date does not establish factual accuracy. AX could assess semantic or interaction-level evidence that these checks leave unresolved, while deterministic results could identify conditions that a conversation export does not reveal. Studies should compare static checks alone, AX alone, and their combination against independent reference assessments and task outcomes, including coverage, diagnostic precision, evaluation cost, and usefulness to developers. Presenting the different sources of evidence separately would also help avoid counting the same problem more than once.

\subsection{Testing the Context-Improvement Feedback Loop}
\label{sec:fw-feedback}

Longitudinal studies are needed to evaluate the proposed feedback loop described in Section~\ref{sec:discussion}. Teams could receive AX assessments with supporting evidence, review the diagnoses, and make targeted changes to shared context. Subsequent work would then test whether those changes reduce recurring deficiencies and improve independently assessed outcomes. Comparison conditions, such as ordinary review without AX feedback, would help distinguish the contribution of the instrument from the effects of additional attention or context maintenance alone.

The evaluation should examine correctness, review and rework, elapsed time, measured cost, and processed token volume as separate outcomes. For software changes, it could also examine maintenance and code retention over a defined period, interpreted alongside defects and legitimate requirements changes. Retention alone would not establish quality. Human outcomes matter as well: does the feedback improve developers' understanding and ability to specify work, and is the review effort worthwhile? Evidence that improvements transfer to later tasks and other team members would be particularly relevant to the proposed cumulative benefit. Higher AX scores alone would not establish that the loop is effective.

\subsection{Security as a Separate Validation Program}

Measuring the Security dimension requires a research program suited to its evidence needs and failure consequences. Initial work should define the assets and policies being protected, the relevant trust boundaries, and the failures the assessment is intended to detect. Evaluation could then combine checks of access and data-handling constraints with adversarial scenarios and observations of tool behavior, where such evidence can appropriately be collected. Some properties may admit deterministic checks; others will require contextual interpretation or targeted testing.

Validation should emphasize missed failures and their severity, rather than relying on average agreement or an undifferentiated accuracy score. Cases with no visible failure should be distinguished from cases with sufficient evidence to support a security judgment. Security findings could contribute to a broader view of context health without being forced into the same ordinal scale as CAFE ratings. Developing and evaluating that complementary approach would extend the practical scope of CAFE(S) while preserving the interpretation of the instrument studied here.

% Complete Conclusion section for all_files/8.Conclusion.tex.
% No new packages or bibliography entries are required.

\section{Conclusion}
\label{sec:conclusion}

As agents take on more development work, understanding their effectiveness requires examining the information environment in which they operate. We introduced Agent Experience (AX), an instrument that operationalizes the Clarity, Actionability, Fidelity, and Efficiency dimensions of CAFE(S) through rubric-guided assessment of session evidence. Inspired by the diagnostic perspective of developer experience, AX produces dimension-specific ratings and supporting explanations to help identify conditions that may impede effective execution. This advances CAFE from a conceptual account of context quality toward an explicit measurement practice.

The three empirical components provide complementary evidence. Reported context problems accompanied lower perceived AI usefulness across development activities. Naturalistic evaluations produced differentiated ratings across sessions and dimensions, while showing that assessment configuration affects the scores. A manual review of a small sample found no problems. In the exploratory scenario study, all eleven defect conditions processed more tokens than the baseline, even though every run produced correct activity-level estimates: context defects appeared as cost before they appeared as error. A diagnostic rule capturing high friction or a decline of at least two rubric categories from the clean reference flagged targeted dimensions at rates from 50\% to 85\% depending on the evaluator model, so evaluator choice mattered as much as the rule. Actionability emerged as the clearest priority: it received the lowest ratings in practice and the hardest for evaluators to detect.

Non-targeted flags most often paired Clarity with Actionability, particularly when Clarity was targeted, which makes the boundary between these two dimensions a priority for calibration. The results provide initial evidence that AX responds to context defects; validating precise localization across tasks and settings and confirming accuracy with independent raters are the next steps.

AX complements checks of information availability, evaluations of models, harnesses, and tools, and independent measures of outcomes and resource use. Its scope is the CAFE evidence available in recorded interactions; Security requires a separate assessment approach. Reviewed AX findings could help teams clarify requirements, correct shared knowledge, and improve the context supplied to subsequent work. Testing that feedback loop is an important next step. This paper supplies a concrete instrument and an empirical account of its behavior, giving teams a structured way to investigate the information conditions on which agentic work depends.

\section*{Conflict of Interest}
All authors are employed by DX. The AX rubric and prompt evaluated in this paper are a proprietary instrument used in DX's Agent experience product and the session transcripts analyzed in Section 3.3 were collected from participants in a pilot of that product.

\section*{Declaration of Generative AI Use}
The authors used OpenAI's ChatGPT and Anthropic's Claude to assist with literature identification and reference checking; manuscript organization, drafting, and revision; and discussion of analytical choices, interpretation, and limitations. ChatGPT and Claude also supported inspection of the supplied study records and development and review of code for recomputing descriptive statistics and producing tables and figures, including checks of calculations and consistency between the recorded evidence and the claims made in the paper. Inputs included prompts, run notes, analysis scripts, recorded scores and resource-use measures, and an example session-transcript export. The use of language models to execute the scenario task and generate AX assessments was part of the research procedure and is described in Methods. The authors retain responsibility for the final analyses, interpretation, references, and manuscript content.

\begin{acks}
We are grateful to all pilot participants and everyone who helped shape this work. In particular, we thank Rudrajit Choudhuri, Autumn Faust, and Scott Carey for their invaluable feedback and support. 
\end{acks}

\bibliographystyle{ACM-Reference-Format}
\bibliography{all_files/bibliography}

\end{document}